# Quantum channel using photon number correlated twin beams

**Yun Zhang*, Katsuyuki Kasai, and Kazuhiro Hayasaka**

Kansai Advanced Research Center, Communications Research Laboratory

588-2 Iwaoka, Nishi-ku, Kobe, 651-2492 Japan

**Abstract**.   We report quantum communications channel using photon number correlated twin beams. The twin beams are generated from a nondegenerate optical parametric oscillator, and the photon number difference is used to encode the signal. The bit error rate of our system will be 0.067 by using the twin beams comparing with 0.217 by using the coherent state as the signal carrier.

* Corresponding author: Yun Zhang

Tel: 81-78-969-2287; Fax: 81-78-969-2219; Email: zhangyun@crl.go.jp

# 1 Introduction

In the past twenty years, a variety of nonclassical states of light has been successfully generated and applied to highly sensitive measurements with sensitivities beyond the shot noise limit (SNL). In recent years, potential application in quantum information has renewed the interest in the generation and application of the nonclassical state. Optical parametric oscillators (OPOs) have turned out to be promising sources for the generation of nonclassical lights. Particularly, it has been experimentally used to generate the quantum correlated twin beams operating above its threshold [1], quadrature squeezed state operating below its threshold [2] and entanglement state with seed injection [3].

It is well known that quantum correlated twin beams would be an elegant means of verifying the foundations of quantum physics. Since the twin beams were first generated experimentally, they have been extensively studied and have been applied to sub-shot-noise [4], quantum nondemolition measurements [5], and most recently FM spectroscopy [6]. The photon-number distribution and joint photon probability distribution of pulse twin beams have been reported [7-9], and recently, the photon-number distribution and joint photon probability distribution of continuous wave twin beams have been investigated by means of direct detection [10,11]. Based on the conditional probability, a bright sub-Poissonian beam from twin beams have also been generated by the technique of conditional measurement performed on continuous variables [12]. In a recent paper, Funk et al. proposed a scheme to realize the quantum key distribution using twin beams [13] in such a scheme the measurement observable is the photon number difference between the signal and idler of outputs from OPOs. Sub-shot-noise quantum correlation between signal and idler gives the necessary sensitivity to eavesdropping that ensures the security of the protocol. The success of these experiments and the potential application scheme in quantum information motivated us to design a quantum communications channel using twin beams. In this paper we present a quantum communications channel using the quantum correlated twin beams. This result shows potential power of Funk and Raymer's method [13] of quantum key distribution using twin beams.

## 2 Quantum communication channel with twin beams

We start by presenting the protocol and deriving the formula for the bit error rate (BER). In order to imitate the scheme of Funk et al, two different polarization bases are used in our experiment. One basis ("V/H basis") is defined by the vertical and horizontal linear polarizations; the other basis ("±45 basis") is defined by the +45 degree linear polarization and -45 degree linear polarization. Alice encodes each bit value in the mean "photon difference number" $\langle n \rangle = \langle n_1 - n_2 \rangle$, where $\langle n_1 \rangle$ ($\langle n_2 \rangle$) is the mean number of photons in the first (second) polarization mode making up a basis. Alice encodes a logical key in either V/H basis or "±45 basis. Bob measures the photon number difference either in the V/H or in the ±45 basis. Thus, the photon number difference in the correct bases, which means Alice and Bob use the same basis, is $n_V - n_H$ in V/H basis and $n_{+45} - n_{-45}$ in ±45 basis respectively. Alice encodes a logical "1" ("0") key bit by setting the mean value of the difference number to be in the correct basis $\langle n \rangle = +N$ ($\langle n \rangle = -N$), where N is a positive number. The photon number difference, N, is small (≤1%) compared with $\langle n_V \rangle$ and $\langle n_H \rangle$, which should contain more than $10^4$ photons on average. After all the key bits have been transmitted, Alice and Bob communicate via a public channel and compare the bases they used for each encoding/measurement. Alice and Bob can separate their communications wrong basis and uncorrected basis dependence on they used the same and difference basis.

For the correct-basis measurement, Bob sets the threshold $N_0$ ( ≥0 ) and constructs his bit sequence by using the following decision:

$$\text{bit value} = \begin{cases} 1 & \text{if } \langle n \rangle > N_0 \\ 0 & \text{if } \langle n \rangle < -N_0 \\ \text{inconclusive} & \text{otherwise,} \end{cases} \qquad (1)$$

where n is the result of Bob's photon number difference measurement. Alice's bit values are determined by the photon number difference: she regards $\langle n \rangle = +N$ as "1" and $\langle n \rangle = -N$ as "0". Due to the tails of the distributions for the two bit values, there is a nonzero probability that a key bit encoded by Alice as a logical 1 (0) would be measured by Bob as a logical 0 (1). Such an error is a bit error (i.e. $1 \leftrightarrow 0$). According to the central limit theorem [14], Bob's photon number difference measurements follow a Gaussian distribution. The distribution can be written as

$$p(n) = \frac{1}{\sqrt{2\pi}\delta} e^{-(n-\langle n \rangle)^2/2\delta^2} + \frac{1}{\sqrt{2\pi}\delta} e^{-(n+\langle n \rangle)^2/2\delta^2}. \qquad (2)$$

The probability has two free parameters, which are the priori selected photon number difference $\langle n \rangle$ and the standard derivation δ of the photon number difference. With this probability, we define the postselection efficiency as the probability that the absolute value of photon number difference $|\langle n \rangle|$ exceeds the threshold $N_0$ in the correct basis. The postselection efficiency is [15]

$$\begin{aligned}P(N_0, \langle n \rangle, \delta) &= \int_{-\infty}^{N_0} p(n)dn + \int_{N_0}^{\infty} p(n)dn \\ &= \frac{1}{2}\{\text{erfc}[\frac{1}{\sqrt{2}\delta}(N_0 - \langle n \rangle)] + \text{erfc}[\frac{1}{\sqrt{2}\delta}(N_0 + \langle n \rangle)]\}\end{aligned}, \qquad (3)$$

where

$$\text{erfc}(z) = \frac{2}{\sqrt{\pi}} \int_{z}^{\infty} e^{-t^2} dt. \qquad (4)$$

For a given $N_0$, the BER can be written as the probability that Bob's measurements have an outcome $\langle n \rangle < N_0$ when Alice has sent the $\langle n \rangle = N$ divided by $P(N_0, \langle n \rangle, \delta)$

$$\begin{aligned}\text{BER} &= \frac{1}{P(N_0,\langle n\rangle,\delta)}\int_{-\infty}^{-N_0} \mathrm{p}\frac{1}{\sqrt{2\pi}\delta}e^{-(n+\langle n\rangle)^2/2\delta^2}dn \\ &= \frac{1}{2P(N_0,\langle n\rangle,\delta)}\mathrm{erfc}[\frac{1}{\sqrt{2}\delta}(N_0+\langle n\rangle)].\end{aligned} \qquad (5)$$

The BER is a function of $N_0$, $\langle n\rangle$ and $\delta$. For a given system, after the parameters $\langle n_1\rangle$, $\langle n_2\rangle$ and $N_0$ have been chosen; the BER thus becomes a function of only the photon number difference standard derivation $\delta$. If one uses a state, in which the standard derivation of photon number difference is less than that of a coherent state, a quantum communications with a lower BER can be realized. Quantum correlated twin beams satisfy this condition. It was demonstrated that the photon number difference between two polarization modes of twin beams is better defined than that of a coherent state with the same total number of photons, i.e., the photon number difference standard derivation, $\delta$, of twin beams is less than that of a coherent state.

## 3 Experimental setup and result

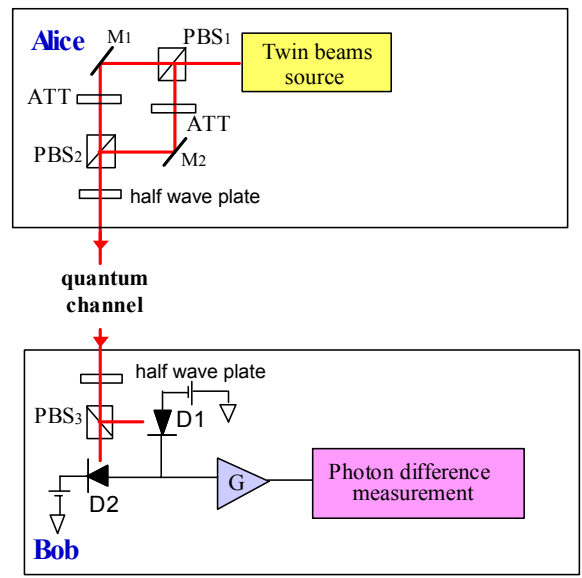

Fig. 1 Schematic of the experimental setup. PBS: polarizing beam splitter; D1 and D2: detector; G: low noise electronic amplifier; M1 and M2: mirror.

Figure 1 shows the experiment setup. At Alice's station, a nondegenerate optical parametric oscillator (NOPO) pumped above its threshold by a diode-pumped continuous wave (CW) frequency-doubled YAG laser emits quantum correlated twin beams. A detailed description of the NOPO can be found in reference [16]. The intensity correlation between the twin beams is reduced to about -5.5±0.3 dB. The orthogonal polarized twin beams, which correspond to V- and H-polarized modes, are separated by a polarizing beam splitter (PBS1). An attenuator (≤1%), which weakly affects the quantum correlation between the two modes of the twin beams, is inserted in the vertical or horizontal light's path. The attenuator generates the mean value of the difference in photon number between the two modes (either N or -N) so that the logical key bit is encoded. The two beams are then combined by another polarizing beam splitter (PBS2). A half-wave plate, which is rotated by an angle of 0° or +45°, is inserted into the beam paths before the twin beams are send to Bob. As the experimental measurement, when the two beams are rotated by an angle of 0° with respect to its axes, the V/H basis is selected and when the polarizations of the two beams is rotated by an angle of 45°, the ±45 basis are selected. At Bob's station, he measures the photon number difference n in either the V/H basis or the ±45 basis, which is chosen at random using the half-wave plate, PBS3, and a pair of balanced high quantum efficiency InGaAs photodiodes (Eptiaxx ETX500). By biasing the two diodes with opposite polarity and summing their output, one obtains the difference in current. The current is detected with a low-noise amplifier (CLC425), which we accurately calibrated. In order to correctly detect the logical key bit, the beams are rotated by an angle of 0° or -45° at Bob's station. In this situation, the vertical and the horizontal polarization mode from the NOPO always beat detector D1 and D2 in the correct basis, respectively. For comparison with a system implementing with a classical coherent state, the coherent light is input onto another port of PBS1 (not shown in Fig. 1). In the experiment, we could block either the coherent light or twin beams to measure the probability distribution in photon number difference of the system. The BER of our system is measured when either the coherent light is blocked or the twin beams are blocked.

We have implemented the same technique as in ref. [10] to measure the photon number difference between two beams at a given Fourier frequency $\Omega$. The difference in photocurrents is amplified and multiplied by a sinusoidal current of frequency $\Omega$ produced by a signal generator, and filtered by a 100 kHz low-pass filter in order to obtain the instantaneous value of the photocurrent Fourier component at frequency $\Omega$, which is then recorded by an A/D board installed in a PC, which also simultaneously records the instantaneous value of the DC photocurrent. In direct detection with a photodetector, assuming that most electrons in the current are photoelectrons, the number of photoelectons measured by the detector is proportional to the photon number of the input state [17]. Since the output from the detector is difference in current, the recorded data are proportional to the difference in photoelectron number, which carries information about the photon number difference of the two beams. Figure 2 shows the photoelctron difference probability distribution of the system implemented with coherent light and twin beams in correct basis and incorrect basis. It was obtained by subdividing about 100 000 experimental data, which were obtained with a demodulation frequency $\Omega$=4MHz, into 100 equal-interval bins. The width of bin, $\delta N$, is about 20. The average photoelectron number of each detector was scaled to be $\langle n_1 \rangle = \langle n_2 \rangle = (4.0 \pm 0.1) \times 10^4$, and the photoelectron difference was fixed at $\langle n \rangle = N = 200$, which is about 0.5% of the total photon number $\langle n_1 \rangle (\langle n_2 \rangle)$. The measured mean and standard deviation are listed in Table I. Theoretical probability distributions from equation with $\langle n \rangle = 200$ and standard derivation $\delta_{twin}$=145±10 for twin beams, which corresponds to a 70% reduction in the noise below the SNL, and $\delta_{coh}$=270±10 for coherent beams are also shown in Fig. 2 (blue and yellow curves). To fit the theoretical prediction, a fixed coefficient was introduced to scale all probabilities of the experiment.

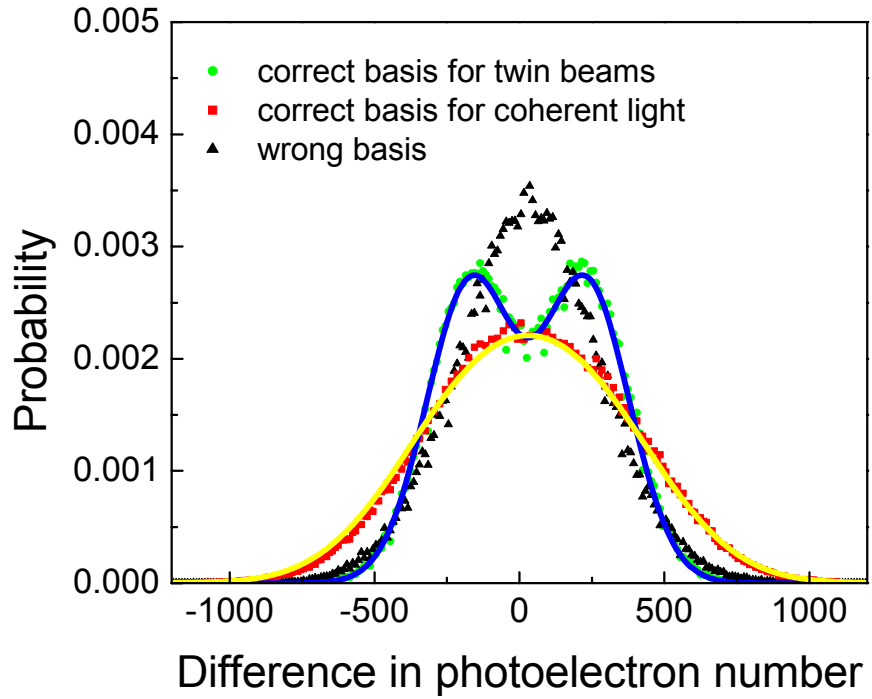

Fig. 2 Measured distributions of photoelectron difference of twin beams and coherent light in correct and incorrect basis. Solid line: theoretical prediction; Symbols: experimental result; Circles: for twin beams in correct basis; Squares: for coherent light in correct basis; Triangles: in wrong basis

Alice and Bob can implement communications by using either a pulse or continuous twin beams source. For the pule source, the photon number difference of the each signal pulses should be measured. For the continuous source, the photon number should be measured in a series of fixed time intervals. To establish the right timing of their recording Alice and Bob have to synchronize their clocks and agree on a set of time intervals $t_k$, in which they subdivide their measurements. Using Eq. (5) and the measured experiment data, the BERs of our system are 0.217 and 0.067 for the coherent light and twin beams, respectively, when we selected $N_0=20$. In a practice implementation, the improvement of the communication can be optimal by selection of the parameters.

We should mention that our quantum communication channel could be used for quantum key distribution as the Funk and Raymer's protocol. The security of the key is ensured by using the quantum correlation between signal and idler of two beams. The quantum correlation between the orthogonal polarization modes is rapidly degraded by any attempt by eavesdropper (Eve) to measure the key. Degradation in the correlation would lead to an increase in the BER that would indicate to Bob and Alice the presence of Eve. In our system the attenuation on one of the twin beams also decrease the degree of correlation between twin beams, but it will not strongly impact the correlations. Because the attenuation is very small ($\leq 1\%$), it can be as the effect as quantum efficiency of detector for the system. Furthermore, the security of the key is ensured by the variation of the BER by any attack. Because the BER is a function of not only δ, but also a priori selected parameter $\langle n \rangle$ and postselected parameter $N_0$, the security of the system also depends on parameters $\langle n \rangle$ and $N_0$. In a practice implementation, the parameters should be determined to ensure that the system provides high security.

Table I. Measured mean and standard derivation of different bases when the system is implemented with coherent state and twin beams.

| Soures | Pre Selection | | photon number difference | |
|---|---|---|---|---|
| | Basis Selection | key | mean | standard deviation |
| Twin Beams | corrected basis | 1 | 200 | 145±10 |
| | corrected basis | 0 | -200 | 145±10 |
| | wrong basis | 1 | 0 | 270±10 |
| | wrong basis | 0 | 0 | 270±10 |
| Coherent Light | corrected basis | 1 | 200 | 270±10 |
| | corrected basis | 0 | -200 | 270±10 |
| | wrong basis | 1 | 0 | 270±10 |
| | wrong basis | 0 | 0 | 270±10 |

## 4 Conclusion

In conclusion, we have presented a new quantum communications channel with photon correlated twin beams. We measured the system's of probability distributions of photon number difference, which is implemented with coherent light or twin beams of -5.5 dB noise reduction. The BERs of the twin beams and coherent light systems were calculated from the experimental data and their comparison indicated the improvement of communication by using the twin beams. This result shows potential power of Funk and Raymer's method of quantum key distribution using twin beams.